\input{aipcheck}

\documentclass[
    ,final            
  ]
  {aipproc}

\layoutstyle{6x9}

\newcommand{\eqr}[1]{Eq.~(\ref{#1})}

\newcommand{\beq}{\begin{equation}}
\newcommand{\eeq}{\end{equation}}
\newcommand{\beqa}{\begin{eqnarray}}
\newcommand{\eeqa}{\end{eqnarray}}
\newcommand{\beqan}{\begin{eqnarray*}}
\newcommand{\eeqan}{\end{eqnarray*}}

\newcommand{\erf}[1]{(\ref{#1})}

\newcommand{\bra}[1]{\langle{#1}|}
\newcommand{\ket}[1]{|{#1}\rangle}

\begin{document}

\title{The Consumption of Reference Resources}


\author{G.~A.~White \ J.~A.~Vaccaro \ H.~M.~Wiseman}{
  address={Center for Quantum Computer Technology, Center for
  Quantum Dynamics, School of Science, Griffith University, Brisbane,
  4111 Australia}
}

\begin{abstract}
Under the operational restriction of the U(1)-superselection rule,
states that contain coherences between eigenstates of particle
number constitute a resource. Such resources can be used to
facilitate operations upon systems that otherwise cannot be
performed. However, the process of doing this consumes reference
resources. We show this explicitly for an example of a unitary
operation that is forbidden by the U(1)-superselection rule.
\end{abstract}

\maketitle


\section{Introduction}
Superselection rules (SSRs) are the constraint that one cannot
observe coherences between eigenstates of a particular quantity
\cite{review}. They were originally discovered by Wight, Wickman and
Wigner in the context of conservation laws \cite{WWW}. However, it
is now customary to think of them as arising from a practical
restriction that one may need to consider due to the presence of a
conservation law, or otherwise \cite{review}. A SSR can in general
be alleviated by an ancillary system that acts as a reference for
the coherences \cite{Suss}. Under a SSR there exists a group
$\{T(g):g\in G\}$ of unitary operators whose effect on the system is
undetectable. The notation we adopt here is that $\{T(g):g\in G\}$
is a representation of the abstract symmetry group $G=\{g\}$.
Consider a system which has been prepared in the state $\rho$.  As
$\rho$ is indistinguishable from $T(g)\rho T^\dagger(g)$ for any
$g\in G$, the {\it effective} state system of the system is given by
the ``twirl'' ${\cal G}_G[\rho]$ of $\rho$ which is defined as [3]
\beq
   {\cal G}_G[\rho]=\frac{1}{|G|}\sum_{g\in G}T(g)\rho T^\dagger(g)\ ,
   \label{twirl definition}
\eeq where $|G|$ is the order of $G$. This indistinguishability
implies that the SSR restricts the allowed operations ${\cal O}$,
including measurements and unitaries, that can be performed upon the
system. The allowed operations are those that are $G$ invariant in
the sense that \cite{wisemanvac}  \beq
      {\cal O} [ \hat{T} (g) \rho \hat{T}^\dagger (g) ]
      = \hat{T} (g) {\cal O} [\rho ] \hat{ T}^\dagger (g)\quad\forall g\in G\ . \label{Ginv}
\eeq

The definition of the twirl in \eqr{twirl definition} is easily
extended to composite systems.  We restrict ourselves to composite
systems where the component subsystems have the same symmetry
described by the symmetry group, $G$. This means that the
representation of the group $G$ for the composite system is
generated by the tensor product of the representations for the
subsystems. Under this restriction the twirl of a composite system
in state $\rho$ is given by \eqr{twirl definition} where the
composite-system operators $T(g)$ have a decomposition of the form
\beq
    T(g)=T_1(g)\otimes T_2(g)\otimes \cdots\quad \forall \ g\in G\ .
\eeq Here $T_i(g)$ acts on the Hilbert space of subsystem labeled
$i$.

\section{Explicit example of a non U(1)-invariant unitary}
 As shown in the previous section, the effect of a
SSR can be to reduce the purity of a particular state. Consider an
arbitrary state, $\rho$, that may be mixed and has a von-Neumann
entropy equal to $S(\rho)$. The von Neumann entropy of the effective
state ${\cal G}\rho$ is greater that that of $\rho$, in general
\cite{wisemanvac,HMW}. The increase in the von Neumann entropy
 is a
measure of the {\it asymmetry}, $A_G(\rho)$, of the state $\rho$
with respect to the symmetry group $G$.  The asymmetry $A_G(\rho)$
of $\rho$ is defined as \cite{vawj} \beq
      A_G(\rho)= S({\cal G}_G[\rho])-S(\rho)
      \label{asymmetry definition}
\eeq where $S(\rho)$ is the von Neumann entropy of $\rho$.
$A_G(\rho)$ has been shown to quantify the ability of an ancilla in
state $\rho$ to act as a reference and partially alleviate the
effects of the $G$ SSR \cite{vawj}.

In general the use of a reference ancilla to perform a non
U(1)-invariant operation on the system may increase the asymmetry of
the system with respect to the U(1) group of phase translations.
Using a reference ancilla repeatedly in this way will reduce the
reference ability of the ancilla. This effect has been studied
explicitly for the case where a reference is used to facilitate a
phase measurement \cite{degrade}. Here we wish to examine this
process for the case of a non U(1)-invariant unitary operation.
 Consider a system that is in the vacuum state \beq
          \ket{\psi}_S = \ket{0}_S \ . \label{qubit}
\eeq The operation $F_{\rm S}$ on the system for which \beq
          F_S (\ket{0}_S )=\frac{1}{\sqrt{2}}(\ket{0}_S +e^{i\theta}\ket{1}_S)
\label{forbidden} \eeq is forbidden under the SSR. However, an
analogous operation can be performed on the coherent subspaces of
the combined reference ancilla and system as follows. We define an
analogous operation on the basis $\{\ket{n}_R\ket{0}_S,
\ket{n-1}_R\ket{1}_S\}$ of the coherent subspaces representing $n$
particles by \beq  F ^{(n)}_{\rm RS} \ket{n}_R\ket{0}_S =
\frac{1}{\sqrt{2}}( \ket{n}_R\ket{0}_S +
          \ket{n-1}_R\ket{1}_S) \label{forb} \eeq for $n=1,2,\ldots,N$. One can perform this operation
with an ancilla say, of the form $\ket{n}$ and produce the the right
hand side of \erf{forb}. However, upon performing the operation one
can easily see that the partial trace over the reference would leave
the system in a maximally mixed state. To minimize this effect use a
phase state of the form \begin{figure} \caption{The consumption of
the reference ancilla is demonstrated by the reduction of the
normalized asymmetry for reference ancilla of $M=5,10,15,20,25$ and
$30$ as a function of the number of times, $\mu$, the reference is
used. Here the normalization, $\eta$, is given by $\log _2[D_R]$.}
\includegraphics[height=.25\textheight]{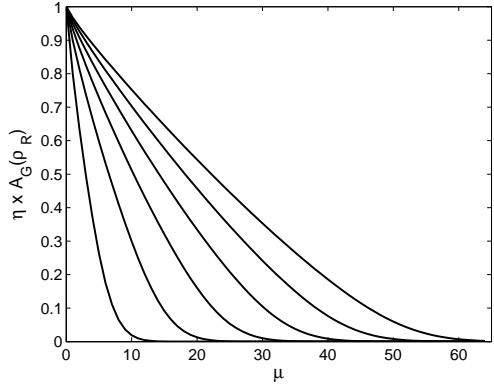}\label{unitary}
\end{figure} \cite{trunc} \beq \ket{\psi}_R =
\frac{1}{\sqrt{N+1}}\sum _{n=0}^N e^{i\theta n}\ket{n}_R \ ,
\label{phase_state} \eeq as the state of the reference ancilla. This
state has the property that it maximizes the asymmetry of the
reference state which has at most $N$ particles. To further simplify
matters we set $\theta = 0$.
 The forbidden operation, ${\cal F}_S$. can be mapped onto a globally G-invariant
operation that acts on the combined state to give \beqa
     \ket{\Psi}_{RS}\equiv F_{\rm RS} \ket\psi_{\rm R}\ket{0}_S &=& \frac{1}{\sqrt{2(N+1)}}
\sum _{n=1}^{N}     \left[  \ket{n}_R\ket{0}_S \right. \nonumber
\\ &+& \left. \ket{n-1}_R\ket{1}_S)\right]+
\frac{1}{\sqrt{N+1}}\ket{0}_R\ket{0}_S \ . \eeqa This state is
entangled. Therefore the reduced state of the reference ancilla is
mixed. Consider a system in the vacuum state $\ket{0}$. We begin
with a reference ancilla in a phase state and perform the forbidden
operation outlined above. The reduced state of the reference is then
given by \beq \rho _R^\prime = Tr_S[\ket{\Psi}\bra{\Psi}] \ . \eeq
The reference ancilla is then used to facilitate the same operation
on another system which is also prepared in the state $\ket{0}$.
This process is repeated $\mu$ times and the resources, i.e. the
asymmetry, of the reference ancilla is consumed as $\mu$ increases.
The consumption of the reference ancilla for reference ancillae of
various size is shown in figure \ref{unitary} as a function of the
number of times, $\mu$, the reference is used to perform the
U(1)-SSR forbidden operation, $F_S$.

\begin{figure}
\caption{The degradation in fidelity of the system state with a
$\ket{+}$ state for $M=5,10,15,20,25$ and $30$ where $\mu$ is the
number of times the reference has been used.}
\includegraphics[height=.25\textheight]{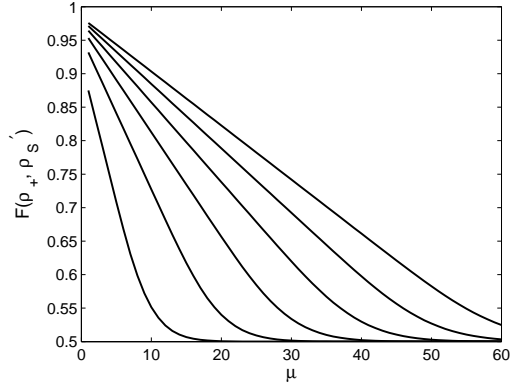}\label{fidelity}
\end{figure}

Consider now the reduced state of the system. Due to the
entanglement between the system and the reference ancilla, the
reduced state of the system is mixed and therefore the fidelity
between the system with the state on the right hand side of
\erf{forbidden} is not perfect. To see this we first define the
fidelity of the operation as \beq F(\rho _S^\prime,\rho _+) = {\rm
Tr}[ \rho _S^\prime \ket{+}\bra{+}] \eeq where \beq \rho _S ^\prime
= {\rm Tr} _R[ F_S \rho _R \ket{0}_S\bra{0}] \ {\rm and} \ \ket{+} =
\frac{1}{\sqrt{2}}(\ket{0}+\ket{1}) \ . \eeq The fidelity the final
state of the system shown in Figure \ref{fidelity}. We see that the
reference's ability to facilitate a non U(1)-invariant unitary
reduces as the reference resources of the reference ancilla are
consumed. The similarity of the trends in figures
\ref{unitary}-\ref{fidelity} show that as the reference show that as
the reference is consumed its ability to facilitate non
U(1)-invariant operations diminishes.

\begin{theacknowledgments}
This work was supported by the Australian Research Council
\end{theacknowledgments}



\bibliographystyle{aipproc}   

\bibliography{sample}

\IfFileExists{\jobname.bbl}{}
 {\typeout{}
  \typeout{******************************************}
  \typeout{** Please run "bibtex \jobname" to optain}
  \typeout{** the bibliography and then re-run LaTeX}
  \typeout{** twice to fix the references!}
  \typeout{******************************************}
  \typeout{}
 }

\end{document}